# An Electrohydrodynamics Model for Non-equilibrium Electron and Phonon Transport in Metal Films after Ultra-short Pulse Laser Heating

Jun Zhou,[1,2] Nianbei Li,[2] and Ronggui Yang[1,♥]

[1]Department of Mechanical Engineering and Materials Science and Engineering Program, University of Colorado, Boulder, CO 80309, USA

[2]Center for Phononics and Thermal Energy Science, School of Physics Science and Engineering, Tongji University, Shanghai 200092, People's Republic of China

**ABSTRACT**

The electrons and phonons in metal films after ultra-short pulse laser heating are in highly non-equilibrium states not only between the electron sub-system and the phonon sub-system but also within the electron sub-system. An electrohydrodynamics model consisting of the balance equations of electron density, energy density of electrons, and energy density of phonons is derived from the coupled non-equilibrium electron and phonon Boltzmann transport equations to study the nonlinear transport phenomena, such as the electron density fluctuation and the transient electrical current in metal films, after ultra-short pulse laser heating. The time-dependent temperature distributions is calculated by the coupled electron and phonon Boltzmann transport equations, the electrohydrodynamics model derived in this work, and the two-temperature model for different laser pulse durations, film thicknesses, and laser fluences. We find that the two-temperature model overestimates the electron temperature at the front surface of the film and underestimates the damage threshold when the nonlinear thermal transport of

---

[♥] Email: Ronggui.Yang@Colorado.Edu

electrons is important. The electrohydrodynamics model proposed in this work could be a more accurate prediction tool to study the non-equilibrium electron phonon transport process than the two-temperature model and it is much easier to be solved than the coupled electron and phonon Boltzmann transport equations.

## I. INTRODUCTION

The study of the interaction of ultra-short pulse laser with materials has a wide range of applications in the fields of fundamental physics of electron-phonon (EP) interaction,[1,2,3] hot electron dynamics,[4,5,6,7,8] superconductivity,[9,10] and laser-material processing.[11,12,13,14,15,16] The energy deposited from ultra-short pulse laser could drive electrons and phonons to a highly nonequilibrium state. For laser-metal interaction, the energy temporarily stored in the electron sub-system is conducted away through two channels: i) electron transport and ii) coupling of electron lattice to the lattice which is then conducted away by phonons.[17,18] The interplay between these two processes determines the evolutions of both electrons and phonons.

The research on this topic can be traced back to as early as 1957. Kaganov *et al.*[17] presented a model for EP energy exchange which is still quite widely used. This phenomenological two-temperature model (TTM) assumes that the sub-systems of electrons and phonons can be described by their own temperatures and both the electron and phonon temperatures are much higher than the Debye temperature. The EP energy exchange rate can then be linearized as $G(T_e - T_p)$ where $T_e$ is the electron temperature, $T_p$ is the phonon temperature, and $G$ is the EP coupling constant. The coupled electron and phonon thermal transport could thus be described by the energy balance equations[18,19]

$$C_e(T_e)\frac{\partial T_e}{\partial t} = -\nabla \cdot \mathbf{J}_{e,E} - G(T_e - T_p) + Q(\mathbf{r},t), \tag{1a}$$

$$C_p(T_p)\frac{\partial T_p}{\partial t} = -\nabla \cdot \mathbf{J}_{p,E} + G(T_e - T_p). \tag{1b}$$

Here $C_e(T_e)$ is the electronic heat capacity, $C_p(T_p)$ is the lattice heat capacity, and $Q(\mathbf{r},t)$ is the energy deposition by the ultra-short pulse laser, $\mathbf{J}_{e,E}$ and $\mathbf{J}_{p,E}$ are the energy current densities of electrons and phonons, respectively. The TTM follows closely the Fourier's law of heat conduction $\mathbf{J}_{e,E} = -\kappa_e(T_e,T_p)\nabla T_e$ and $\mathbf{J}_{p,E} = -\kappa_p(T_p)\nabla T_p$, where $\kappa_e(T_e,T_p)$ is the electronic thermal conductivity and $\kappa_p(T_p)$ is the lattice (phononic) thermal conductivity.

Although the TTM has been widely used by many researchers, there are quite some limitations of TTM to describe the ultra-short pulse laser-material interaction, especially with the recent advances of femtosecond and attosecond lasers. One of the obvious drawback is not considering the transient electrical current flow due to non-equilibrium electrons:

(1) $\mathbf{J}_{e,E} = -\kappa_e(T_e,T_p)\nabla T_e$ used in TTM is valid only when the electrical current and the chemical potential gradient can be ignored.

(2) The divergence of energy current density of electrons $-\nabla \cdot \mathbf{J}_{e,E}$ consists of not only Fourier's heat conduction but also the electron density accumulation, the Thomson effect, and the Joule heat generation.[20] The TTM considers the heat conduction and neglects the other mechanisms.

(3) The approximation $\nabla \cdot [\kappa_e(T_e,T_p)\nabla T_e] \approx \kappa_e(T_e,T_p)\nabla^2 T_e$ is used in the TTM. It is valid only when the thermal conductivity are temperature-independent. In addition, the electronic heat capacity is usually chosen to be proportional to the electron temperature which is valid only when the temperature is lower than Fermi temperature.

Currently, there have been a few works which address the non-equilibrium transport that considers transient electrical current and chemical potential gradient. Chen et al.[21] added an additional conservation equation of electron momentum which is derived from the BTE into the TTM to study the effect of electronic kinetic pressure on the electron and phonon temperature evolutions. As a result, they obtain significantly different electron and phonon temperature response from the TTM. The larger laser fluence is and shorter laser pulse duration is, the larger the difference between their model and the TTM is.

In this paper, we derive an electrohydrodynamics (EHD) model based on the coupled electron and phonon BTE to investigate the electron and phonon transport in metals after the ultra-short pulse laser heating. In this model, the energy current is driven by both the temperature gradient and the chemical potential gradient. The contributions to $-\nabla \cdot \mathbf{J}_{e,E}$ from the heat conduction, the electron density accumulation, the Thomson effect, and the Joule heat generation are considered. The paper is organized as follows. After a brief introduction of the BTE in Sec. IIA, we derive the EHD model in detail in Sec. IIB. In Sec. III, the results of the calculation of the electron density fluctuation, the transient electrical current, the evolutions of electron temperature and phonon temperature, and the laser fluence damage threshold are presented. Section IV concludes this work.

## II. THEORETICAL MODELS

Figure 1 shows the coupled non-equilibrium electron and phonon transport mechanisms in a metal film after ultra-short pulse laser heating. The surface of the film is in *x-y* plane (perpendicular to *z*-direction) and the film thickness is *L*. The photon energy of the ultra-short pulse laser can be absorbed by electrons near the front surface ($x = 0$) of the film. The intensity of absorption exponentially decays with the depth where $\delta$ is the optical absorption depth. Highly non-equilibrium states between the electron sub-system and the phonon sub-system are induced because the electrons are excited by ultra-short pulse laser to higher energy states. The highly non-equilibrium electrons will dissipate their energy through two channels: i) electron transport and ii) coupling of electron lattice to the lattice which is then conducted away by phonons. Diffusive reflection boundary conditions are used in this study by assuming that electrons and phononsare reflected back randomly when they reach the front surface or the rear surface of the film.

**A. Coupled electron and phonon BTE model**

We can write the coupled electron and phonon BTE as follows: [22]

$$\frac{\partial f_{\mathbf{k}}(\mathbf{r},t)}{\partial t} + \mathbf{v}_{e,\mathbf{k}} \cdot \nabla f_{\mathbf{k}}(\mathbf{r},t) - \frac{e}{\hbar}\mathbf{E} \cdot \nabla_{\mathbf{k}} f_{\mathbf{k}}(\mathbf{r},t) = \left.\frac{\partial f_{\mathbf{k}}(\mathbf{r},t)}{\partial t}\right|_{c} + s(\mathbf{r},\mathbf{k},t), \quad (2a)$$

$$\frac{\partial n_{\mathbf{q}}(\mathbf{r},t)}{\partial t} + \mathbf{v}_{p} \cdot \nabla n_{\mathbf{q}}(\mathbf{r},t) = \left.\frac{\partial n_{\mathbf{q}}(\mathbf{r},t)}{\partial t}\right|_{c}, \quad (2b)$$

where $f_{\mathbf{k}}(\mathbf{r},t)$ is the electron distribution function in the phase space with wave vector **k** and position $\mathbf{r} = (x, y, z)$ at time $t$, $n_{\mathbf{q}}(\mathbf{r},t)$ is the phonon distribution function in the phase space with wave vector **q** [23] and position **r** at time $t$. $\nabla_{\mathbf{k}}$ is the wave vector gradient operator, and **E** is the external electric field. $\mathbf{v}_{e,\mathbf{k}} = \frac{1}{\hbar}\nabla_{\mathbf{k}}\varepsilon_{k}$ is the electron velocity, where $\varepsilon_{k}$ is the kinetic energy of electron and $\hbar$ is the Planck constant. $\mathbf{v}_{p}$ is the phonon velocity. $s(\mathbf{r},\mathbf{k},t)$ is the source term due to laser heating.[24] $\partial f_\mathbf{k}(\mathbf{r},t)/\partial t |_c$ is the electron collision term which includes the EP scattering term $\partial f_\mathbf{k}(\mathbf{r},t)/\partial t |_{ep}$ and the electron-electron scattering term $\partial f_\mathbf{k}(\mathbf{r},t)/\partial t |_{ee}$. $\partial n_\mathbf{q}(\mathbf{r},t)/\partial t |_c$ is the phonon collision term which includes the EP scattering term $\partial n_\mathbf{q}(\mathbf{r},t)/\partial t |_{ep}$ and the phonon-phonon scattering term $\partial n_\mathbf{q}(\mathbf{r},t)/\partial t |_{pp}$.

The EP scattering terms in Eq. (2) are calculated from a microscopic approach:[9]

$$\left.\frac{\partial f_\mathbf{k}}{\partial t}\right|_{ep} = -\frac{2\pi}{\hbar}\sum_\mathbf{q}|M_\mathbf{q}|^2 \{\delta(\varepsilon_k - \varepsilon_{k'} - \hbar\omega_q)[n_\mathbf{q}(f_\mathbf{k} - f_{\mathbf{k}'}) + f_\mathbf{k}(1 - f_{\mathbf{k}'})] \\ - \delta(\varepsilon_k - \varepsilon_{k'} + \hbar\omega_q)[n_\mathbf{q}(f_{\mathbf{k}'} - f_\mathbf{k}) + f_{\mathbf{k}'}(1 - f_\mathbf{k})]\}, \quad (3a)$$

$$\left.\frac{\partial n_\mathbf{q}}{\partial t}\right|_{ep} = -\frac{4\pi}{\hbar}\sum_\mathbf{k}|M_\mathbf{q}|^2 f_\mathbf{k}(1 - f_{\mathbf{k}'})[n_\mathbf{q}\delta(\varepsilon_k - \varepsilon_{k'} + \hbar\omega_q) \\ - (n_\mathbf{q} + 1)\delta(\varepsilon_k - \varepsilon_{k'} - \hbar\omega_q)] \quad (3b)$$

We drop $(\mathbf{r},t)$ in the rest of this paper for simplification. Here $M_q$ is the EP scattering matrix elements, $\hbar\omega_q$ is the energy of phonons whose wave vector equals to the wave vector change of electron after the scattering $\mathbf{q} = \mathbf{k} - \mathbf{k}'$. Delta functions denote the energy conservation of the scattering processes. Electron spin degeneracy has been taken into account by the factor 2 in Eq. (3b).

To calculate the EP scattering terms in Eq. (3) in metal without doing the summation over three dimensional momentum space, the Eliashberg function [25] $\alpha^2 F(\omega)$ is introduced similar as shown in Ref. [9]. We then have

$$\left.\frac{\partial f_\mathbf{k}}{\partial t}\right|_{ep} = -2\pi\int d\omega \alpha_k^2 F(\omega) \int d\varepsilon' \{\delta(\varepsilon_k - \varepsilon' - \hbar\omega)[n_{0,\omega}(f_{0,\varepsilon_k} - f_{0,\varepsilon'}) + f_{0,\varepsilon_k}(1 - f_{0,\varepsilon'})] \\ - \delta(\varepsilon_k - \varepsilon' + \hbar\omega)[n_{0,\omega}(f_{0,\varepsilon'} - f_{0,\varepsilon_k}) + f_{0,\varepsilon'}(1 - f_{0,\varepsilon_k})]\}, \quad (4a)$$

$$\left.\frac{\partial n_{\mathbf{q}}}{\partial t}\right|_{ep} = -2\pi\Xi(\mu)\int d\varepsilon \int d\varepsilon' \frac{\alpha^2 F(\omega)}{F(\omega)}(f_{0,\varepsilon} - f_{0,\varepsilon'})$$
$$\left[n_{0,\omega}\delta(\varepsilon - \varepsilon' + \hbar\omega) - (n_{0,\omega} + 1)\delta(\varepsilon - \varepsilon' - \hbar\omega)\right]\bigg|_{\omega=\omega_q},$$
(4b)

where $\Xi(\mu)$ is the electron density states (DOS) at the chemical potential and $F(\omega)$ is the phonon DOS. $f_{0,\varepsilon_k}$ is the equilibrium Fermi-Dirac distribution of electrons and $n_{0,\omega}$ is the equilibrium Bose-Einstein distribution of phonons.

In non-equilibrium systems, the electron-electron scattering and the phonon-phonon scattering drive the non-equilibrium electron and phonon sub-systems towards to equilibrium ones, respectively. Therefore, we assume $\partial f_{\mathbf{k}}/\partial t|_{ee} = -\left[f_{\mathbf{k}} - f_{0,\varepsilon_k}\right]/\tau_{ee}(k)$ [24] and $\partial n_{\mathbf{q}}/\partial t|_{pp} = -\left[n_{\mathbf{q}} - n_{0,\omega_q}\right]/\tau_{pp}(q)$ using relaxation time approximation where $\tau_{ee}(k)$ is the electron-electron relaxation time and $\tau_{pp}(q)$ is the phonon-phonon relaxation time. In this work, $\tau_{ee}(k)$ and $\tau_{pp}(q)$ are assumed to be constants so that we could focus on the EP scattering on the energy transfer pathway.

With the scattering terms explicitly written as above, the coupled electron and phonon BTE model shown in Eq. (2) can be solved iteratively for both directions ($k_z > 0$ and $k_z < 0$) by using finite difference method with a numerical scheme similar to that described in Ref. [26]. In our calculation, the Eliashberg function is taken from Ref. [27] and the phonon DOS is taken from Ref. [28]. Both the front surface and rear surface are assumed to be rough so that the equilibrium distribution function due to diffusive reflection of electrons and phonons can be written at the boundaries:[29]

$$f_{\mathbf{k}}(z=0,t) = f_{0,\varepsilon_k}(z=0,t),\ f_{\mathbf{k}}(z=L,t) = f_{0,\varepsilon_k}(z=L,t), \quad (5a)$$

$$n_{\mathbf{q}}(z=0,t) = n_{0,\omega_q}(z=0,t),\ n_{\mathbf{q}}(z=L,t) = n_{0,\omega_q}(z=L,t). \quad (5b)$$

Table I lists all the parameters used in the calculation of BTE.

**B. Electrohydrodynamics (EHD) Model**

The coupled electron and phonon BTE is a set of multi-variable integral-differential equations in the phase space where the scattering integral is very complicated to be solved.[30] A convenient approach to simplify the problem is to use balance equations, instead of directly solving the BTE to obtain the distribution function of electrons and phonons, where the macroscopic variables such as electron density, electron temperature, and phonon temperature are calculated. The computational cost is remarkably reduced in spite of the losing of some detailed information in distribution functions. Here, we write a set of EHD equations with the electron density and energy density of electrons as macroscopic variables by multiplying $1/V$ and $\varepsilon_k/V$ in Eq. (2a) and then summing over $\mathbf{k}$, where $V$ is the unit volume.[22]

$$\frac{\partial N}{\partial t} = -\nabla \cdot \mathbf{J}_{e,n}, \tag{6a}$$

$$\frac{\partial U_e}{\partial t} = -\nabla \cdot \mathbf{J}_{e,E} - e\mathbf{E} \cdot \mathbf{J}_{e,n} - G(T_e - T_p) + Q(\mathbf{r},t). \tag{6b}$$

Here $N = 2\sum_{\mathbf{k}} f_{\mathbf{k}}$ is the local electron density and $\mathbf{J}_{e,n} = 2\sum_{\mathbf{k}} \mathbf{v}_{e,\mathbf{k}} f_{\mathbf{k}}$ is the particle current density of electrons ($-e\mathbf{J}_{e,n}$ is the electrical current density). $U_e = 2\sum_{\mathbf{k}} \varepsilon_k f_{\mathbf{k}}$ is the energy density of electrons and $\mathbf{J}_{e,E} = 2\sum_{\mathbf{k}} \mathbf{v}_{e,\mathbf{k}} \varepsilon_k f_{\mathbf{k}}$ is the energy current density of electrons. The factor 2 comes from electron spin degeneracy. $\mathbf{J}_p = \sum_{\mathbf{q}} \mathbf{v}_p \hbar \omega_q n_q$ is the thermal current density carried by phonons. We focus our study on noble metals, the photoelectric effect and inter-band transition which excites *d*- or *f*-electrons to be itinerant electrons in transition metals are not considered. Therefore, $2\sum_{\mathbf{k}} s(\mathbf{r},\mathbf{k},t) = 0$ since there is no net electron generated or annihilated in electron intra-band transition due to photon absorption as shown in Fig. 1. The electron energy source term is $Q = 2\sum_{\mathbf{k}} \varepsilon_k s(\mathbf{r},\mathbf{k},t)$ which is assumed to be independent on the temperature.[19] The balance equation which is the same as Eq. 1(b) can also be obtained by multiplying $\hbar\omega_q / V$ in Eq. (2b) and then summing over $\mathbf{q}$.

Under the free electron approximation, the left side of Eq. (6b) can be written as:

$$\frac{\partial U_e}{\partial t} = \frac{\partial U_e}{\partial \mu}\frac{\partial \mu}{\partial t} + \frac{\partial U_e}{\partial T_e}\frac{\partial T_e}{\partial t} = \frac{\partial U_e}{\partial \mu}\left(\frac{\partial N}{\partial \mu}\right)^{-1}\frac{\partial N}{\partial t} + C_e(T_e)\frac{\partial T_e}{\partial t}, \qquad (7)$$

where the electronic heat capacity is defined as $C_e(T_e) = \frac{\partial U_e}{\partial T_e} - \frac{\partial U_e}{\partial \mu}\left(\frac{\partial N}{\partial \mu}\right)^{-1}\frac{\partial N}{\partial T_e}$ which can be calculated as

$$C_e(T_e) = \gamma T_e\left[1 - \frac{3}{10}\left(\frac{k_B T_e}{\mu_0}\right)^2 + O\left(\frac{k_B T_e}{\mu_0}\right)^4\right], \qquad (8)$$

where $\mu_0$ is the chemical potential at zero temperature.[31] See Appendix A for detailed derivations.

$\mathbf{J}_{e,n}$ and $\mathbf{J}_{e,E}$ in Eqs. (6a) and (6b) can be obtained by solving the BTE under the relaxation time approximation[22] as shown in Appendix B for details:

$$\mathbf{J}_{e,n} = -\frac{\sigma}{e}\mathbf{E} - D_n \nabla N - S_n \nabla T_e, \qquad (9a)$$

$$\mathbf{J}_{e,E} = -\left(-T_e\sigma\alpha + \frac{\mu\sigma}{e}\right)\mathbf{E} - D_E\nabla N - S_E\nabla T_e. \tag{9b}$$

Here $\sigma$ is the electrical conductivity, $D_n = \frac{\sigma}{e^2}\frac{2\mu_0}{3N_0}$ is the diffusion coefficient of electron, $S_n = -\frac{\sigma\alpha}{e}\frac{2\lambda-1}{2\lambda}$ is the Soret coefficient, $\alpha$ is the Seebeck coefficient, $D_E = -\frac{T_e\sigma\alpha}{e}\frac{2\mu_0}{3N_0} + \mu D_n$, and $S_E = \kappa_e^0 - T_e\sigma\alpha^2\frac{1}{2\lambda} + \mu S_n$, where $\kappa_e^0 = \kappa_e + T_e\sigma\alpha^2$. Here $N_0$ is the electron density without fluctuation. $\lambda = -\alpha\left(\frac{\pi^2 k_B^2 T_e}{3e\mu_0}\right)^{-1}$ is almost a constant for given metal.[32]

Then $-\nabla\cdot\mathbf{J}_{e,n}$ and $-\nabla\cdot\mathbf{J}_{e,E}$ in Eqs. (6a) and (6b) can be calculated as follows (see detailed derivation in Appendix C):

$$-\nabla\cdot\mathbf{J}_{e,n} \approx D_n\nabla^2 N + S_n\nabla^2 T_e - \frac{D_n}{3N_0}(\nabla N)^2 - \frac{2S'_n}{T_e}(\nabla T_e)^2 - \frac{2D_n}{T_e}(\nabla N\cdot\nabla T), \tag{10a}$$

$$-\nabla\cdot\mathbf{J}_{e,E} \approx D_E\nabla^2 N + S_E\nabla^2 T_e - \frac{D_E}{3N_0}(\nabla N)^2 - \frac{2S'_E}{T_e}(\nabla T_e)^2 - \frac{2D_E}{T_e}(\nabla N\cdot\nabla T_e). \tag{10b}$$

Here $S'_n = -\frac{\sigma\alpha}{e}\frac{4\lambda-1}{4\lambda}$ and $S'_E = \kappa_e^0 - T_e\sigma\alpha^2\frac{1}{4\lambda} + \mu S'_n$.

In the rest of this work, we do not consider the external electric field for simplification. By substituting Eq. (10) into Eq. (6), the EHD model can finally be written as (see Appendix D for details):

$$\frac{\partial N}{\partial t} = D_n\nabla^2 N + S_n\nabla^2 T_e - \frac{D_n}{3N_0}(\nabla N)^2 - \frac{2S'_n}{T_e}(\nabla T_e)^2 - \frac{2D_n}{T_e}(\nabla N\cdot\nabla T), \tag{11a}$$

$$C_e(T_e)\frac{\partial T_e}{\partial t} = D\nabla^2 N + S\nabla^2 T_e - \frac{D}{3N_0}(\nabla N)^2 - \frac{2S'}{T_e}(\nabla T_e)^2 - \frac{2D}{T_e}(\nabla N \cdot \nabla T_e)$$ (11b)
$$- G(T_e - T_p) + Q,$$

$$C_p(T_p)\frac{\partial T_p}{\partial t} = \kappa_p(T_p)\nabla^2 T_p + \frac{\partial \kappa_p(T_p)}{\partial T_p}(\nabla T_p)^2 + G(T_e - T_p),$$ (11c)

where $D = -\frac{T_e \sigma \alpha}{e}\frac{2\mu_0}{3N_0}\frac{2\lambda-1}{2\lambda}$, $S = \kappa_e^0 - T_e\sigma\alpha^2\frac{4\lambda-1}{4\lambda^2}$, and $S' = \kappa_e^0 - T_e\sigma\alpha^2\frac{6\lambda-1}{8\lambda^2}$.

The models for the temperature dependence of the transport coefficients are chosen as follows in our calculation: $\sigma = \sigma_{rt}\frac{T_{rt}}{T_p}$, $\alpha = \alpha_{rt}\frac{T_e}{T_{rt}}$,[32] $\kappa_e^0 = \kappa_{e,rt}^0 \frac{T_e}{T_p}$,[18] where $\sigma_{rt}$, $\alpha_{rt}$, and $\kappa_{e,rt}^0$ are the transport coefficients when both electron temperature and phonon temperature are at room temperature $T_{rt} = 300\text{K}$. The electron energy source term is written as[19]

$$Q(\mathbf{r},t) = 0.94(1-R)I\frac{1}{\tau_p \delta}\exp\left[-\frac{z}{\delta} - 2.77\frac{t^2}{\tau_p^2}\right],$$ (12)

where $R$ is the reflectivity, $I$ is the fluence of laser pulse, and $\tau_p$ is the laser pulse time duration. The time zero is noted as the center of laser pulse, $t = 0$.

The thermal-insulation boundary conditions for temperature and zero electrical current boundary conditions are used:

$$\nabla T_e\big|_{z=0} = \nabla T_e\big|_{z=L} = 0, \quad \nabla T_p\big|_{z=0} = \nabla T_p\big|_{z=L} = 0, \quad \nabla N_e\big|_{z=0} = \nabla N_e\big|_{z=L} = 0.$$ (13)

The parameters used in the EHD model calculation in gold thin film are listed in Table I.

## C. Comparison between TTM and EHD Model

By comparing TTM in Eq. (1) with EHD model in Eq. (11), the major differences between the proposed EHD model and the TTM are as follows:

(1). The addition of the balance equation on electron density in the EHD model (Eq. (11a)) enables us to study the electron density fluctuation and transient electrical current induced by ultra-short pulse laser heating. The chemical potential gradient $\nabla \mu$ is considered as the driving force for current flow in the EHD model, which is not even considered in the TTM model. A complete energy current density of electron $\mathbf{J}_{e,E} = \mathbf{J}_{e,Q} + \mu \mathbf{J}_{e,n}$ is considered in the EHD while $\mu \mathbf{J}_{e,n}$ is not included in TTM.

(2). The divergence of energy current density of electron in the presence of electrical current is[20]

$$-\nabla \cdot \mathbf{J}_{e,E} = \nabla \cdot [\kappa_e(T_e,T_p)\nabla T_e] - \mu \nabla \cdot \mathbf{J}_{e,n} - \xi \mathbf{J}_{e,n} \cdot \nabla T_e + \frac{1}{\sigma}[e\mathbf{J}_{e,n}]^2, \quad (14)$$

where $\xi$ is the Thomson coefficient. Comparing to the TTM model, the EHD model considers the electron accumulation, the Thomson effect, and the Joule heat generation through the last three terms on the right side of Eq. (14)

(3). The divergence of heat current of electrons is

$$\nabla \cdot [\kappa_e(T_e,T_p)\nabla T_e] = \kappa_e(T_e,T_p)\nabla^2 T_e + \left[\frac{\partial \kappa_e(T_e,T_p)}{\partial T_e}\right](\nabla T_e)^2 + \left[\frac{\partial \kappa_e(T_e,T_p)}{\partial T_p}\right]\nabla T_p \cdot \nabla T_e, \quad (15)$$

where the first term on the right side in describes the linear thermal transport due to the electron temperature gradient while the last two terms are additional nonlinear thermal transport terms. The TTM model only explicitly considers the linear transport term.[19] Empirically, the nonlinear terms can be calculated when one has an explicit analytical temperature dependence of $\kappa_e(T_e,T_p)$. Anisimov and Rathfeld[33] showed that $\kappa_e \propto T_e/T_p$ when $T_e \ll T_F$ and $\kappa_e \propto T_e^{5/2}$ when $T_e \sim T_F$, where $T_F$ is the Fermi temperature of material (usually over $10^4$ K in metals). A more complicated analytical expression of $\kappa_e$ is given by Refs. [34,35]. The question here is whether it is correct to directly substitute the above temperature dependence of $\kappa_e$ into Eq. (15)? A revisit is needed from more basic approaches such as the BTE. We assume $\kappa_e \propto T_e/T_p$ as an example and set the Seebeck coefficient $\alpha = 0$ which means $D = 0$ and $S = S' = \kappa_e^0 = \kappa_e$, then the divergence of heat current in Eq. (11b) becomes

$$\kappa_e(T_e, T_p)\nabla^2 T_e - \frac{2\kappa_e(T_e, T_p)}{T_e}(\nabla T_e)^2. \tag{16}$$

In contrast, the divergence of heat current in Eq. (1a) in the TTM can be rewritten as:

$$\kappa_e(T_e, T_p)\nabla^2 T_e + \frac{\kappa_e(T_e, T_p)}{T_e}(\nabla T_e)^2 - \frac{\kappa_e(T_e, T_p)}{T_p}\nabla T_p \cdot \nabla T_e. \tag{17}$$

The coefficients of linear term $\nabla^2 T_e$ in Eq. (16) and Eq. (17) are the same, and the coefficient of nonlinear term $(\nabla T_e)^2$ in Eq. (16) is $-2\kappa_e/T_e$ which is different from $\kappa_e/T_e$ in Eq. (17), and there is one more term $-\frac{\kappa_e}{T_p}\nabla T_p \cdot \nabla T_e$ in Eq. (17). The difference between the EHD model and the TTM can be neglected only when $(\nabla T_e)^2/T_e \ll \nabla^2 T_e$ and $(\nabla T_p \cdot \nabla T_e)/T_p \ll \nabla^2 T_e$. When $(\nabla T_e)^2/T_e \sim \nabla^2 T_e$ and $(\nabla T_p \cdot \nabla T_e)/T_p \sim \nabla^2 T_e$ one have to directly solve the BTE or solve the EHD model instead.

## III. RESULTS AND DISCUSSIONS

In the following, we compare the calculation results from the coupled electron and phonon BTE model, the proposed EHD model and the classical TTM for the coupled electron and phonon transport in gold thin films after ultra-short pulse heating, as show in Fig. 1.

**A. Electron density fluctuation and transient electrical current**

We first compare the evolution of electron temperature at the front surface for a 50nm-thick gold film after an ultra-short pulse laser heating with fluence $I = 1\text{mJ/cm}^2$ and pulse duration $\tau_p = 96\text{fs}$ calculated from the BTE model, the EHD model, and the TTM. The EP coupling constant is chosen to be $G = 2.1 \times 10^{16} \text{W/m}^3\text{K}$ throughout this paper which is obtained by fitting the experimental data measured by Brorson *et al.* in Ref. [8], which is close to the values in literature, $2.2 \times 10^{16} \text{W/m}^3\text{K}$ - $4 \times 10^{16} \text{W/m}^3\text{K}$.**Error! Bookmark not defined.** As shown in Fig. 2(a), the electron temperature at the front surface calculated from the EHD model is in good agreement with the prediction from the BTE while the TTM overestimates the electron temperature.

The large temperature gradient due to the non-equilibrium ultra-short pulse laser heating can induce electron density redistribution and transient electrical current due to electron diffusion and thermoelectric effect. Such electron density fluctuation and transient electrical current after ultra-short pulse laser heating can be calculated by the BTE model and the EHD model while the Fourier-like TTM model is not able to. Figure 2(b) shows the electron density fluctuation in gold film at different time delays, *t*=0.05ps, 0.1ps, and 0.2ps. At the beginning, the amplitude of the electron density fluctuation is on the order of $10^{24}/\text{m}^3$, which is about 1% of the electron density near Fermi surface $N_0 k_B T / \mu_0$. Figure 2(c) shows the particle current density of electrons in gold film at different time delays, $t$=0ps, 0.1ps, 0.2ps. At the beginning, a negative particle current is noted due to flow of particles towards the front surface as a result of thermoelectric effect. However, the particle current density becomes positive, i.e., flow away from the front surface at $t$=0.1ps, due to the diffusion of electrons driven by electron density gradient as shown in Fig. 2(b). The maximum amplitude of the particle current density is about $10^{30}/\text{sm}^2$ which results an electrical current density on the order of $10^{11} \text{A/m}^2$. After $t$=0.2ps, the particle current density becomes negligible. In our calculation, the Seebeck coefficient of gold is positive[36] which results in a negative Soret coefficient ($S_n < 0$) in Eq. (9a). This negative Soret coefficient results in the flow of electrons towards the front surface driven by temperature gradient. In other metals with negative Seebeck coefficient and positive Soret coefficient (for example, Al, Pb *et al.*), the electrons would flow away from the front surface driven by temperature gradient.

**B. Dependence of temperature evolution on laser pulse duration, laser fluence and film thickness**

The time scale that the laser-deposited energy is conducted away from the front surface is on the order of sub-picosecond. When the laser pulse time duration is smaller than this time scale, most energy absorbed from laser is stored at the front surface and the interior of the film is not heated up yet at the beginning of the time evolution. This results in an abrupt temperature drop near the front surface. In other words, $\nabla T_e$ is very large and $(\nabla T_e)^2 / T_e \sim \nabla^2 T_e$ is satisfied which means that the nonlinear thermal transport could be important as we discussed in Sec. IIC. Figure 2(a), Fig. 3(a), and Fig. 3(b) show the evolution of electron temperature at the front surface of the film with thickness $L = 50\text{nm}$ for different laser pulse time durations, $\tau_p = 96\text{fs}$, 150fs, and 200fs, when the laser fluence is $I = 1\text{mJ/cm}^2$. For all these three time durations, the electron temperature calculated from the EHD model is closer to that calculated from BTE while the TTM overestimate the electron temperature. The maximum electron temperatures for the BTE and the EHD model are almost the same when $\tau_p = 96\text{fs}$. The calculated maximum electron temperatures are 634K from the BTE, 660K from the EHD model, and 686K from the TTM when the pulse duration is $\tau_p = 150\text{fs}$. When the duration time increases to 200fs, the maximum electron temperatures become 621K from the BTE, 653K from the EHD model, and 672K from the TTM. The reason is that $(\nabla T_e)^2 / T_e \sim \nabla^2 T_e$ for shorter time duration and the nonlinear thermal transport must be considered.

When the electrons are heated by the laser pulse, they would transport inside the film with the Fermi velocity $v_F$ (about $1.4 \times 10^6 \text{m/s}$ in gold), and then be reflected back by the rear surface. These hot electrons return to the front surface after a time interval, $\sim 2L/v_F$. When this time interval is smaller than the laser pulse duration, the temperature evolution at the front surface can be affected by the boundary reflection from the rear surface which results in highly nonlinear transport. Therefore, the thickness of film is another important parameter in our study. Figures 3(c) and 3(d) show the evolution of electron temperature at front surface of the film for different thicknesses, $L = 40\text{nm}$ and 60nm, when the laser fluence is $I = 1\text{mJ/cm}^2$ and the laser pulse duration is $\tau_p = 96\text{fs}$. Figure 2(a) also shows the results with $L = 50\text{nm}$ while other conditions are the same. For all these three film thickness, the electron temperature calculated from the EHD model is closer to that from the BTE, and the TTM overestimates the electron temperature. The calculated maximum electron temperatures are 696K from the BTE, 723K from the EHD model, and 750K from the TTM when $L = 40\text{nm}$. When the film thickness increases to $50\text{nm}$, the electron temperature calculated from the EHD model is almost the same as that from the BTE. When the film thickness is $60\text{nm}$, the electron temperature calculated from the EHD model is slightly smaller than that from the BTE.

Larger laser fluence results in higher electron temperature at the front surface and the electron distribution is farther away from the equilibrium distribution. Therefore, the electron transport is highly nonlinear for large laser fluence. Figures 4(a) and 4(b) show the evolution of electron temperature at front surface of the film for different laser fluences, $I = 5\text{mJ/cm}^2$ and $10\text{mJ/cm}^2$ when the film thickness is $L = 50\text{nm}$ and the laser pulse time duration is $\tau_p = 96\text{fs}$, respectively. Figure 2(a) also shows the results with $I = 1\text{mJ/cm}^2$ when other conditions are the same. For $I = 1\text{mJ/cm}^2$ and $5\text{mJ/cm}^2$, the electron temperature calculated from the EHD model is close to that calculated from the BTE for all the time delays considered. The TTM overestimates the electron temperature. For higher laser fluence $I = 10\text{mJ/cm}^2$, both the EHD model and the TTM underestimate the electron temperature in comparison with the BTE at the beginning of the time evolution (time delay<1.5ps). The reason is that both the EHD model and the TTM assume that the electron distribution is slightly deviated from the Fermi-Dirac distribution which does not hold at the beginning of the time evolution for high laser fluence. When time delay>1.5ps, the electron temperatures calculated from the EHD model and the BTE become getting closer since the electron-electron scattering drives the electron sub-system from non-equilibrium distribution to equilibrium one which is called thermalization process.<sup>Error! Bookmark not defined.</sup>

From Fig. 3 and Fig. 4, we find that the EHD model is more accurate than TTM in thinner film and shorter laser pulse time duration because of the contribution from transient electrical current and nonlinear thermal transport. When the thermalization process is important for higher laser fluence, both the EHD model and the TTM underestimate the electron temperature at the beginning of the time evolution.

**C. Damage thresholds**

We now study the laser damage threshold calculated from the EHD model and the TTM with high laser fluences. We do not show the results from the BTE calculations due to the computational cost. The BTE model calculates the distribution functions in wave vector and position phase-space. Larger laser fluence requires larger truncation wave vector that significantly increases the computational cost. Figure 5(a) shows that the time evolutions of both electron temperature and phonon temperature at the front surface of the film calculated from the EHD model slightly differ from that calculated from the TTM when $L = 25\text{nm}$. Therefore, the converged temperature (electron temperature becomes equal to phonon temperature) calculated from the EHD model is also slightly smaller than that from TTM. When the film thickness increases to 50nm, the maximum electron temperature calculated from the EHD model could be 800K lower than that from TTM and the converged temperature calculated from EHD model is 100K higher than that from TTM as shown in Fig. 5(b). Overall, the EHD model gives a lower converged temperature than the TTM, which means that more energy at the front surface is conducted away from the surface.

When the laser fluence is large enough, the phonon temperature at the front surface can reach the melting point of the metal, which is 1337K for gold. Such laser fluence is called damage fluence threshold. As we shown in Fig. 5, the converged temperature depends on the film thickness. Therefore, the damage fluence threshold is also a function of film thickness. Figure 6 shows the calculated damage fluence thresholds from the EHD model and the TTM for an ultra-short pulse laser for different film thickness with a laser pulse duration $\tau_p = 96\text{fs}$. As shown in Figs. 5 and 6, the TTM overestimates the converged temperature at front surface of the film and the difference of this overestimation increases with the film thickness. Therefore, the EHD model gives a larger damage fluence threshold in comparison with the TTM. For example, when the film thickness is smaller than 25nm, the damage fluence thresholds calculated from the EHD model and the TTM are both around $125\text{mJ/cm}^2$. When the film thickness is increased to 60nm, the damage fluence threshold calculated from the EHD model is $375\text{mJ/cm}^2$ which is much larger than that calculated from the TTM, $236\text{mJ/cm}^2$.

## IV. CONCLUSIONS

In summary, we established an EHD model which consists of the balance equations of electron density, energy density of electrons, and energy density of phonons to investigate the coupled non-equilibrium electron and phonon transport in thin metal films after the ultra-short pulse laser heating. The non-equilibrium transport of electrons such as the electron density fluctuation and transient electrical current are considered in the EHD model beyond the phenomenological TTM. By comparing the calculated results from the BTE for different laser pulse time durations, film thicknesses, and laser fluences, we found that the EHD model is more accurate than the TTM which usually overestimates the surface electron temperature when the nonlinear thermal transport in electron transport is important. We find that shorter laser pulse duration and thinner metal film result in stronger nonlinear transport. This model gives a more accurate tool to study the non-equilibrium coupled electron and phonon transport process than the TTM and it is much easier to be solved than the BTE.

**Acknowledgements**: J. Z. would like to thank Dr. J. L. Cheng and Dr. Y. Y. Wang for valuable discussions. R. Y. would like to acknowledge the support from AFOSR (Grant No. FA9550-11-1-0109) and NSF (Grant No. 0846561). J. Z. and B. L. are supported by the National Natural Science Foundation of China grant no. 11334007. J. Z. is also supported by the program for New Century Excellent Talents in Universities grant no. NCET-13-0431.

# APPENDIX A. ELECTRONIC HEAT CAPACITY

We calculate the temperature-dependent electronic heat capacity by using the Sommerfeld expansion method under the free electron approximation and local equilibrium approximation, the local electron density and energy density can be written as:

$$N = \chi \int_0^\infty d\varepsilon_k \frac{\varepsilon_k^{1/2}}{\exp[(\varepsilon_k - \mu)/k_B T_e + 1]} = \chi(k_B T_e)^{3/2} F_{1/2}(\eta), \qquad (A1a)$$

$$U_e = \chi \int_0^\infty d\varepsilon_k \frac{\varepsilon_k^{3/2}}{\exp[(\varepsilon_k - \mu)/k_B T_e + 1]} = \chi(k_B T_e)^{5/2} F_{3/2}(\eta), \qquad (A1b)$$

where $\eta = \mu/k_B T_e$, $F_j(\eta)$ is the dimensionless integral $F_j(\eta) = (k_B T_e)^{-j-1} \int_0^\infty d\varepsilon_k \varepsilon_k^j /[\exp(\varepsilon_k/k_B T_e - \eta) + 1]$, the electron density of states is $\Xi(\varepsilon_k) = \chi \varepsilon_k^{1/2}$, $\chi = \frac{V}{2\pi^2}\left(\frac{2m}{\hbar^2}\right)^{3/2}$, $V$ is the unit volume, $m$ is the electron mass, and $j = -1/2, 1/2, 3/2$.

The derivatives in Eq. (7) can be calculated as

$$\frac{\partial N}{\partial \mu} = \frac{1}{2}\chi(k_B T_e)^{1/2} F_{-1/2}(\eta), \qquad (A2a)$$

$$\frac{\partial N}{\partial T_e} = \frac{1}{2T_e}\chi(k_B T_e)^{3/2}\left[3F_{1/2}(\eta) - \frac{\mu}{k_B T_e} F_{-1/2}(\eta)\right], \qquad (A2b)$$

$$\frac{\partial U_e}{\partial \mu} = \frac{3}{2}\chi(k_B T_e)^{3/2} F_{1/2}(\eta), \qquad (A2c)$$

$$\frac{\partial U_e}{\partial T_e} = \frac{1}{2T_e}\chi(k_B T_e)^{5/2}\left[5F_{3/2}(\eta) - \frac{\mu}{k_B T_e} 3F_{1/2}(\eta)\right]. \qquad (A2d)$$

Here we have used $\frac{dF_{1/2}(\eta)}{d\eta} = \frac{1}{2}F_{-1/2}(\eta)$, $\frac{dF_{3/2}(\eta)}{d\eta} = \frac{3}{2}F_{1/2}(\eta)$. The Taylor expansion of $F_j(\eta)$ can be found in Ref. [31]

$$F_{3/2}(\eta) = \frac{2}{5}\eta^{5/2}\left[1 + \frac{5\pi^2}{8}\eta^{-2} - \frac{7\pi^4}{384}\eta^{-4} + O(\eta^{-6})\right], \quad \text{(A3a)}$$

$$F_{1/2}(\eta) = \frac{2}{3}\eta^{3/2}\left[1 + \frac{\pi^2}{8}\eta^{-2} + \frac{7\pi^4}{640}\eta^{-4} + O(\eta^{-6})\right], \quad \text{(A3b)}$$

$$F_{-1/2}(\eta) = 2\eta^{1/2}\left[1 - \frac{\pi^2}{24}\eta^{-2} - \frac{7\pi^2}{384}\eta^{-4} + O(\eta^{-6})\right]. \quad \text{(A3c)}$$

By ignoring the high order terms, the electronic heat capacity defined for Eq. (7) can be written as

$$\begin{aligned}
C_e(T_e) &= \frac{\partial U_e}{\partial T_e} - \frac{\partial U_e}{\partial \mu}\left(\frac{\partial N}{\partial \mu}\right)^{-1}\frac{\partial N}{\partial T_e} \\
&= \frac{1}{2T_e}\chi(k_B T_e)^{5/2}\left[5F_{3/2}(\eta) - 9F_{1/2}^2(\eta)/F_{-1/2}(\eta)\right] \\
&\approx \frac{1}{3}\chi\pi^2 k_B^2 \mu^{1/2} T_e\left[1 - \frac{31\pi^2}{120}\left(\frac{k_B T_e}{\mu}\right)^2\right] \\
&\approx \gamma T_e\left[1 - \frac{3\pi^2}{10}\left(\frac{k_B T_e}{\mu_0}\right)^2\right].
\end{aligned} \quad \text{(A4)}$$

Here, we use the relation $\mu(T_e) \approx \mu_0\left[1 - \frac{\pi^2}{12}\left(\frac{k_B T_e}{\mu_0}\right)^2\right]$ and replace $\frac{k_B T_e}{\mu}$ [31] in the square bracket by $\frac{k_B T_e}{\mu_0}$ where the higher order terms are ignored. $\gamma = \frac{1}{3}\chi\pi^2 k_B^2 \mu_0^{1/2} = \frac{N_0}{2\mu_0}\pi^2 k_B^2$ is the ideal heat capacity constant.

# APPENDIX B. PARTICLE CURRENT DENSITY AND ENERGY CURRENT DENSITY OF ELECTRONS

The particle current density and energy current density of electrons are calculated by solving electron BTE under relaxation time approximation. The distribution function in the BTE can be written into two parts $f_\mathbf{k} = f_{0,\mathbf{k}} + f_{1,\mathbf{k}}$,[22] where $f_{0,\mathbf{k}} = \dfrac{1}{e^{(\varepsilon_k - \mu)/k_B T_e} + 1}$ is the equilibrium Fermi-Dirac distribution. The non-equilibrium part driven by thermodynamic force along z-direction can be written as:

$$f_{1,\mathbf{k}} = eE_z \frac{\hbar k_z \tau_k}{m} \frac{\partial f_{0,\mathbf{k}}}{\partial \varepsilon_k} - \frac{\hbar k_z \tau_k}{m} \frac{\partial f_{0,\mathbf{k}}}{\partial z}, \tag{B1}$$

with $f_{1,\mathbf{k}}$ under the relaxation time approximation. The particle current density and the energy current density of electrons alone z-direction are defined as

$$J_{e,n,z} = \sum_\mathbf{k} v_{e,\mathbf{k},z} f_{1,\mathbf{k}}, \tag{B2a}$$

$$J_{e,E,z} = \sum_\mathbf{k} v_{e,\mathbf{k},z} \varepsilon_k f_{1,\mathbf{k}}, \tag{B2b}$$

We replace $v_z^2$ by $2\varepsilon_k/3m$ using isotropic condition, and substitute Eq. (B1) into Eq. (B2).

$$\begin{aligned} J_{e,n,z} &= \frac{2e}{3m} \sum_\mathbf{k} \tau_k \varepsilon_k \left[ \frac{\partial f_{0,k}}{\partial \varepsilon_k} E_z - \frac{1}{e} \frac{\partial f_{0,k}}{\partial z} \right] \\ &= -\frac{N}{e} \langle \tau_k \varepsilon_k^{3/2} \rangle \left( eE_z + \frac{\partial \mu}{\partial z} \right) - \frac{N}{eT_e} \langle \tau_k \varepsilon_k^{3/2} (\varepsilon_k - \mu) \rangle \frac{\partial T_e}{\partial z}, \end{aligned} \tag{B3a}$$

$$\begin{aligned} J_{e,E,z} &= \frac{2e}{3m} \sum_\mathbf{k} \tau_k \varepsilon_k^2 \left[ \frac{\partial f_{0,k}}{\partial \varepsilon_k} E_z - \frac{1}{e} \frac{\partial f_{0,k}}{\partial z} \right] \\ &= -\frac{N}{e} \langle \tau_k \varepsilon_k^{5/2} \rangle \left( eE_z + \frac{\partial \mu}{\partial z} \right) - \frac{N}{eT_e} \langle \tau_k \varepsilon_k^{5/2} (\varepsilon_k - \mu) \rangle \frac{\partial T_e}{\partial z}, \end{aligned} \tag{B3b}$$

where the relation $\frac{\partial f_{0,k}}{\partial z} = -\frac{\partial f_{0,k}}{\partial \varepsilon_k}\left[\frac{\partial \mu}{\partial z} + \frac{\varepsilon - \mu}{T_e}\frac{\partial T_e}{\partial z}\right]$ is used for the derivation. The bracket $\langle A(\varepsilon_k)\rangle$ denotes the integration of arbitrary function $A(\varepsilon_k)$

$$\langle A(\varepsilon_k)\rangle = -\frac{2e}{3mN}\chi\int_0^\infty d\varepsilon_k A(\varepsilon_k)\frac{\partial f_{0,k}}{\partial \varepsilon_k}, \tag{B4}$$

The derivative of chemical potential $\frac{\partial \mu}{\partial z}$ can be transformed to the derivative of electron density $\frac{\partial N}{\partial z}$ and electron temperature $\frac{\partial T_e}{\partial z}$ as follows:

$$\begin{aligned}\frac{\partial \mu}{\partial z} &= \left(\frac{\partial N}{\partial \mu}\right)^{-1}\frac{\partial N}{\partial z} - \frac{\partial N}{\partial T_e}\left(\frac{\partial N}{\partial \mu}\right)^{-1}\frac{\partial T_e}{\partial z}\\ &= \frac{2}{\chi(k_B T_e)^{1/2}F_{-1/2}(\eta)}\frac{\partial N}{\partial z} - \left[\frac{3k_B F_{1/2}(\eta)}{F_{-1/2}(\eta)} - \frac{\mu}{T_e}\right]\frac{\partial T_e}{\partial z}\\ &\approx \frac{1}{\chi\mu^{1/2}}\frac{\partial N}{\partial z} - \frac{\pi^2 k_B^2 T_e}{6\mu}\frac{\partial T_e}{\partial z}\end{aligned} \tag{B5}$$

Then Eqs. (B3a) and (B3b) can be rewritten by substituting Eq. (B5) into them

$$\begin{aligned}J_{e,n,z} &= -\frac{N}{e}\langle \tau_k \varepsilon_k^{3/2}\rangle eE_z - \frac{N}{e\chi\mu^{1/2}}\langle \tau_k \varepsilon_k^{3/2}\rangle\frac{\partial N}{\partial z}\\ &\quad -\frac{N}{eT_e}\left[-\frac{\pi^2(k_B T_e)^2}{6\mu}\langle \tau_k \varepsilon_k^{3/2}\rangle + \langle \tau_k \varepsilon_k^{3/2}(\varepsilon_k - \mu)\rangle\right]\frac{\partial T_e}{\partial z}\\ &= -\frac{\sigma}{e}E_z - D_n\frac{\partial N}{\partial z} - S_n\frac{\partial T_e}{\partial z},\end{aligned} \tag{B6a}$$

$$\begin{aligned}J_{e,E,z} &= -\frac{N}{e}\langle \tau_k \varepsilon_k^{5/2}\rangle eE_z - \frac{N}{e\chi\mu^{1/2}}\langle \tau_k \varepsilon_k^{5/2}\rangle\frac{\partial N}{\partial z}\\ &\quad -\frac{N}{eT_e}\left[-\frac{\pi^2(k_B T_e)^2}{6\mu}\langle \tau_k \varepsilon_k^{5/2}\rangle + \langle \tau_k \varepsilon_k^{5/2}(\varepsilon_k - \mu)\rangle\right]\frac{\partial T_e}{\partial z}\\ &= -\left(-T_e\sigma\alpha + \frac{\mu\sigma}{e}\right)E_z - D_E\frac{\partial N}{\partial z} - S_E\frac{\partial T_e}{\partial z}.\end{aligned} \tag{B6b}$$

The transport coefficients are defined as follows: the electrical conductivity is $\sigma = Ne\langle\tau_k\varepsilon_k^{3/2}\rangle$, the diffusion constant of electron $D_n = \frac{\sigma}{e^2}\frac{1}{\chi\mu^{1/2}} \approx \frac{\sigma}{e^2}\frac{2\mu_0}{3N_0}$, the Soret coefficient is $S_n = -\frac{\sigma\alpha}{e}\frac{2\lambda-1}{2\lambda}$, the Seebeck coefficient is $\alpha = -\frac{1}{eT_e}\frac{\langle\tau_k\varepsilon_k^{3/2}(\varepsilon_k-\mu)\rangle}{\langle\tau_k\varepsilon_k^{3/2}\rangle}$, $D_E = -\frac{T_e\sigma\alpha}{e}\frac{2\mu_0}{3N_0} + \mu D_n$, and $S_E = \kappa_e^0 - T_e\sigma\alpha^2\frac{1}{2\lambda} + \mu S_n$. The electronic thermal conductivity is $\kappa_e = \kappa_e^0 - T_e\sigma\alpha^2$ where $\kappa_e^0 = \frac{N}{eT_e}\langle\tau_k\varepsilon_k^{3/2}(\varepsilon_k-\mu)^2\rangle$. The relations $\alpha = -\frac{\pi^2 k_B^2 T_e}{3e\mu_0}\lambda$, $N_0 = \frac{2}{3}\chi\mu_0^{3/2}$, and $\mu \approx \mu_0$ are also used[31] where $\lambda$ is a material parameter.[32] Mathematical tricks $\langle\tau_k\varepsilon_k^{5/2}\rangle = \langle\tau_k\varepsilon_k^{3/2}(\varepsilon_k-\mu)\rangle + \mu\langle\tau_k\varepsilon_k^{3/2}\rangle$ and $\langle\tau_k\varepsilon_k^{5/2}(\varepsilon_k-\mu)\rangle = \langle\tau_k\varepsilon_k^{3/2}(\varepsilon_k-\mu)^2\rangle + \mu\langle\tau_k\varepsilon_k^{3/2}(\varepsilon_k-\mu)\rangle$ are used in the calculation.

The current components in *x* and *y* direction can be calculated in the same way. We can then extend Eqs. (B6a) and (B6b) to three dimension equations

$$\mathbf{J}_{e,n} = -\frac{\sigma}{e}\mathbf{E} - D_n\nabla N - S_n\nabla T_e, \tag{B7a}$$

$$\mathbf{J}_{e,E} = -\left(-T_e\sigma\alpha + \frac{\mu\sigma}{e}\right)\mathbf{E} - D_E\nabla N - S_E\nabla T_e. \tag{B7b}$$

## APPENDIX C. DIVERGENCES OF ELECTRON DENSITY AND ENERGY DENSITY OF ELECTRONS

The divergences of electron density and energy density of electrons can be calculated as:

$$-\nabla \cdot \mathbf{J}_{e,n} = -\frac{\partial}{\partial z} J_{e,n,z} = -\sum_{\mathbf{k}} v_{e,\mathbf{k},z} \frac{\partial}{\partial z} f_{1,\mathbf{k}}, \tag{C1a}$$

$$-\nabla \cdot \mathbf{J}_{e,E} = -\frac{\partial}{\partial z} J_{e,E,z} = -\sum_{\mathbf{k}} v_{e,\mathbf{k},z} \varepsilon_k \frac{\partial}{\partial z} f_{1,\mathbf{k}}, \tag{C1b}$$

when the thermodynamic forces such as temperature gradient and chemical potential gradient are alone $z$-direction, i.e. $-\frac{\partial}{\partial x} J_{e,n,x} = -\frac{\partial}{\partial y} J_{e,n,y} = 0$ and $-\frac{\partial}{\partial x} J_{e,E,x} = -\frac{\partial}{\partial y} J_{e,E,y} = 0$.

We assume that only the equilibrium distribution function in $f_{1,\mathbf{k}}$ vary with position and the external electric field is absent. Eqs. (C1a) and (C1b) become

$$\begin{aligned}
-\frac{\partial}{\partial z} J_{e,n,z} &= -\frac{2e}{3m} \sum_{\mathbf{k}} \tau_k \varepsilon_k \left[ -\frac{1}{e} \frac{\partial^2 f_{0,k}}{\partial z^2} \right] \\
&\approx \frac{N}{e} \langle \tau_k \varepsilon_k^{3/2} \rangle \frac{\partial^2 \mu}{\partial z^2} + \frac{N}{eT_e} \langle \tau_k \varepsilon_k^{3/2} (\varepsilon_k - \mu) \rangle \frac{\partial^2 T_e}{\partial z^2} \\
&- \frac{2N}{eT_e} \langle \tau_k \varepsilon_k^{3/2} \rangle \frac{\partial \mu}{\partial z} \frac{\partial T_e}{\partial z} - \frac{2N}{eT_e} \langle \tau_k \varepsilon_k^{3/2} (\varepsilon_k - \mu) \rangle \left( \frac{\partial T_e}{\partial z} \right)^2,
\end{aligned} \tag{C2a}$$

$$\begin{aligned}
-\frac{\partial}{\partial z} J_{e,E,z} &= -\frac{2e}{3m} \sum_{\mathbf{k}} \tau_k \varepsilon_k^2 \left[ -\frac{1}{e} \frac{\partial^2 f_{0,k}}{\partial z^2} \right] \\
&\approx \frac{N}{e} \langle \tau_k \varepsilon_k^{5/2} \rangle \frac{\partial^2 \mu}{\partial z^2} + \frac{N}{eT_e} \langle \tau_k \varepsilon_k^{5/2} (\varepsilon_k - \mu) \rangle \frac{\partial^2 T_e}{\partial z^2} \\
&- \frac{2N}{eT_e} \langle \tau_k \varepsilon_k^{5/2} \rangle \frac{\partial \mu}{\partial z} \frac{\partial T_e}{\partial z} - \frac{2N}{eT_e} \langle \tau_k \varepsilon_k^{5/2} (\varepsilon_k - \mu) \rangle \left( \frac{\partial T_e}{\partial z} \right)^2.
\end{aligned} \tag{C2b}$$

Here the second order derivative of distribution function is

$$\begin{aligned}
\frac{\partial^2 f_{0,k}}{\partial z^2} &= -\frac{\partial f_{0,k}}{\partial \varepsilon_k} \left[ \frac{1}{k_B T_e} \frac{e^{(\varepsilon_k - \mu)/k_B T_e} - 1}{e^{(\varepsilon_k - \mu)/k_B T_e} + 1} \right] \left( \frac{\partial \mu}{\partial z} + \frac{\varepsilon_k - \mu}{T_e} \frac{\partial T_e}{\partial z} \right)^2 \\
&+ \frac{\partial f_{0,k}}{\partial \varepsilon_k} \frac{2}{T_e} \frac{\partial T_e}{\partial z} \left( \frac{\partial \mu}{\partial z} + \frac{\varepsilon_k - \mu}{T_e} \frac{\partial T_e}{\partial z} \right) - \frac{\partial f_{0,k}}{\partial \varepsilon_k} \left( \frac{\partial^2 \mu}{\partial z^2} + \frac{\varepsilon_k - \mu}{T_e} \frac{\partial^2 T_e}{\partial z^2} \right).
\end{aligned} \tag{C3}$$

The first term on the right in Eq. (C3) is symmetric on the chemical potential, its contribution to $-\frac{\partial}{\partial z}J_{e,n,z}$ and $-\frac{\partial}{\partial z}J_{e,E,z}$ is negligible. That is why we use "≈" in Eqs. (C2a) and (C2b). $\frac{\partial^2 \mu}{\partial z^2}$ in Eqs. (C2a) and (C2b) can be transformed to

$$\frac{\partial^2 \mu}{\partial z^2} = \left(\frac{\partial N}{\partial \mu}\right)^{-1}\left[\frac{\partial^2 N}{\partial z^2} - \frac{\partial N}{\partial T_e}\frac{\partial^2 T_e}{\partial z^2} - \frac{\partial}{\partial z}\left(\frac{\partial N}{\partial \mu}\right)\frac{\partial \mu}{\partial z} - \frac{\partial}{\partial z}\left(\frac{\partial N}{\partial T_e}\right)\frac{\partial T_e}{\partial z}\right]. \tag{C4}$$

And $\frac{\partial \mu}{\partial z}$ in Eqs. (C2a), (C2b), and (C4) can be transformed to $\frac{\partial N}{\partial z}$ and $\frac{\partial T_e}{\partial z}$ using Eq. (B5). Then Eqs. (C2a) and (C2b) are rewritten as

$$\begin{aligned}-\frac{\partial}{\partial z}J_{e,n,z} &\approx \frac{N}{e\chi\mu^{1/2}}\langle\tau_k\varepsilon_k^{3/2}\rangle\frac{\partial^2 N}{\partial z^2} + \frac{N}{eT_e}\left[-\frac{\pi^2(k_BT_e)^2}{6\mu}\langle\tau_k\varepsilon_k^{3/2}\rangle + \langle\tau_k\varepsilon_k^{3/2}(\varepsilon_k-\mu)\rangle\right]\frac{\partial^2 T_e}{\partial z^2}\\ &\quad -\frac{N}{2e\chi^2\mu^2}\langle\tau_k\varepsilon_k^{3/2}\rangle\left(\frac{\partial N}{\partial z}\right)^2 - \left[\frac{2N}{eT_e^2}\langle\tau_k\varepsilon_k^{3/2}(\varepsilon_k-\mu)\rangle - \frac{\pi^2 k_B^2}{6\mu}\langle\tau_k\varepsilon_k^{3/2}\rangle\right]\left(\frac{\partial T_e}{\partial z}\right)^2 \\ &\quad -\frac{2N}{eT_e\chi\mu^{1/2}}\langle\tau_k\varepsilon_k^{3/2}\rangle\frac{\partial N}{\partial z}\frac{\partial T_e}{\partial z}\\ &= D_n\frac{\partial^2 N}{\partial z^2} + S_n\frac{\partial^2 T_e}{\partial z^2} - \frac{D_n}{3N_0}\left(\frac{\partial N}{\partial z}\right)^2 - \frac{2S'_n}{T_e}\left(\frac{\partial T_e}{\partial z}\right)^2 - \frac{2D_n}{T_e}\frac{\partial N}{\partial z}\frac{\partial T_e}{\partial z},\end{aligned} \tag{C5a}$$

$$\begin{aligned}-\frac{\partial}{\partial z}J_{e,E,z} &\approx \frac{N}{e\chi\mu^{1/2}}\langle\tau_k\varepsilon_k^{5/2}\rangle\frac{\partial^2 N}{\partial z^2} + \frac{N}{eT_e}\left[-\frac{\pi^2(k_BT_e)^2}{6\mu}\langle\tau_k\varepsilon_k^{5/2}\rangle + \langle\tau_k\varepsilon_k^{5/2}(\varepsilon_k-\mu)\rangle\right]\frac{\partial^2 T_e}{\partial z^2}\\ &\quad -\frac{N}{2e\chi^2\mu^2}\langle\tau_k\varepsilon_k^{5/2}\rangle\left(\frac{\partial N}{\partial z}\right)^2 - \left[\frac{2N}{eT_e^2}\langle\tau_k\varepsilon_k^{5/2}(\varepsilon_k-\mu)\rangle - \frac{\pi^2 k_B^2}{6\mu}\langle\tau_k\varepsilon_k^{5/2}\rangle\right]\left(\frac{\partial T_e}{\partial z}\right)^2 \\ &\quad -\frac{2N}{eT_e\chi\mu^{1/2}}\langle\tau_k\varepsilon_k^{5/2}\rangle\frac{\partial N}{\partial z}\frac{\partial T_e}{\partial z}\\ &= D_E\frac{\partial^2 N}{\partial z^2} + S_E\frac{\partial^2 T_e}{\partial z^2} - \frac{D_E}{3N_0}\left(\frac{\partial N}{\partial z}\right)^2 - \frac{2S'_E}{T_e}\left(\frac{\partial T_e}{\partial z}\right)^2 - \frac{2D_E}{T_e}\frac{\partial N}{\partial z}\frac{\partial T_e}{\partial z},\end{aligned} \tag{C5b}$$

where $S'_n = -\frac{\sigma\alpha}{e}\frac{4\lambda-1}{4\lambda}$ and $S'_E = \kappa_e^0 - T_e\sigma\alpha^2\frac{1}{4\lambda} + \mu S'_n$.

Eqs. (C5a) and (C5b) can be extended to three dimension form

$$-\nabla \cdot \mathbf{J}_{e,n} \approx D_n \nabla^2 N + S_n \nabla^2 T_e - \frac{D_n}{3N_0}(\nabla N)^2 - \frac{2S'_n}{T_e}(\nabla T_e)^2 - \frac{2D_n}{T_e}(\nabla N \cdot \nabla T),  \quad (\text{C6a})$$

$$-\nabla \cdot \mathbf{J}_{e,E} \approx D_E \nabla^2 N + S_E \nabla^2 T_e - \frac{D_E}{3N_0}(\nabla N)^2 - \frac{2S'_E}{T_e}(\nabla T_e)^2 - \frac{2D_E}{T_e}(\nabla N \cdot \nabla T_e). \quad (\text{C6b})$$

**APPENDIX D. BALANCE EQUATION OF ENERGY DENSITY OF ELECTRONS**

Using the relation in Eq. (7) $\frac{\partial U_e}{\partial t} = \frac{\partial U_e}{\partial \mu}\left(\frac{\partial N}{\partial \mu}\right)^{-1}\frac{\partial N}{\partial t} + C_e(T_e)\frac{\partial T_e}{\partial t}$, the balance equation Eq. (9b) in the absence of external electric field becomes

$$\frac{\partial U_e}{\partial t} = \frac{\partial U_e}{\partial \mu}\left(\frac{\partial N}{\partial \mu}\right)^{-1}\frac{\partial N}{\partial t} + C_e(T_e)\frac{\partial T_e}{\partial t} = -\nabla \cdot \mathbf{J}_{e,E} - G(T_e - T_p) + Q(\mathbf{r},t). \quad (\text{D1})$$

Eq. (D1) can be rewritten as

$$C_e(T_e)\frac{\partial T_e}{\partial t} = -\nabla \cdot \mathbf{J}_{e,E} - \frac{\partial U_e}{\partial \mu}\left(\frac{\partial N}{\partial \mu}\right)^{-1}\frac{\partial N}{\partial t} - G(T_e - T_p) + Q(\mathbf{r},t). \quad (\text{D2})$$

Using the divergences of electron density and energy density of electrons obtained in Eqs. (C6a) and (C6b), the first two terms on the right side of Eq. (D2) are

$$-\nabla \cdot \mathbf{J}_{e,E} - \frac{\partial U_e}{\partial \mu}\left(\frac{\partial N}{\partial \mu}\right)^{-1}\frac{\partial N}{\partial t} \approx -\nabla \cdot \mathbf{J}_{e,E} + \mu\left[1 + \frac{\pi^2 (k_B T_e)^2}{6\mu^2}\right]\nabla \cdot \mathbf{J}_{e,n}$$

$$= D\nabla^2 N + S\nabla^2 T_e - \frac{D}{3N_0}(\nabla N)^2 - \frac{2S'}{T_e}(\nabla T_e)^2 - \frac{2D}{T_e}(\nabla N \cdot \nabla T_e), \quad (\text{D3})$$

where $D = -\frac{T_e \sigma \alpha}{e}\frac{2\mu_0}{3N_0}\frac{2\lambda-1}{2\lambda}$, $S = \kappa_e^0 - T_e \sigma \alpha^2 \frac{4\lambda-1}{4\lambda^2}$, and $S' = \kappa_e^0 - T_e \sigma \alpha^2 \frac{6\lambda-1}{8\lambda^2}$. The relations $\frac{\partial N}{\partial t} = -\nabla \cdot \mathbf{J}_{e,n}$ and $\frac{\partial U_e}{\partial \mu}\left(\frac{\partial N}{\partial \mu}\right)^{-1} = 3k_B T_e \frac{F_{1/2}(\eta)}{F_{-1/2}(\eta)} \approx \mu\left[1 + \frac{\pi^2 (k_B T_e)^2}{6\mu^2}\right]$ are used.

Combining Eqs. (D2) and (D3), the balance equation of energy density of electrons is finally written as:

$$C_e(T_e)\frac{\partial T_e}{\partial t} = D\nabla^2 N + S\nabla^2 T_e - \frac{D}{3N_0}(\nabla N)^2 - \frac{2S'}{T_e}(\nabla T_e)^2 - \frac{2D}{T_e}(\nabla N \cdot \nabla T_e)$$
$$- G(T_e - T_p) + Q(\mathbf{r},t). \quad (D4)$$

Table I. Parameters used in the calculation of the BTE model and the EHD model.

| | | | |
|---|---|---|---|
| $R^a$ | 0.93 | $\delta^a$ | 15.3 nm |
| $\sigma_{rt}^b$ | $4.52\times10^7\,/\Omega\text{m}$ | $\alpha_{rt}^c$ | 1.94μV/K |
| $\kappa_{e,rt}^{0\,a}$ | 315 W/mK | $\lambda^c$ | -1.48 |
| $\mu_0^b$ | 5.5 eV | $N_0^b$ | $5.86\times10^{28}\,/\text{m}^3$ |
| $\gamma^a$ | 62.9 J/m³K² | $C_p^a$ | 70 J/m³K |
| $v_p^b$ | 3240 m/s | $\tau_{ee}^d$ | 0.04ps |
| $\tau_{pp}^d$ | 0.08ps | | |

[a] Ref. [19]; [b] [28]; [c] [32]; [d] Ref. [37]

**Figure Captions**

FIG. 1 (Color online) Schematic diagram of the coupled non-equilibrium electron and phonon transport in metal film after ultra-short pulse laser heating. The film thickness is $L$ and the surface of the metal film is in *x-y* plane. The low energy electrons (black dot) near the front surface are excited to be higher energy ones (red dot) after absorbing a photon. The absorption intensity exponentially decays with the radiation penetration depth $\delta$. The hot electrons release energy through two channels: transport to the rear surface and colliding with the lattice to emit phonons, i.e. EP scattering. When the electrons reach both the front and rear surfaces, they are scattered back randomly if a rough surface is assumed. In this work, diffusive reflection boundary conditions of electrons and phonons are used.

FIG. 2 (Color online) (a) Time-dependent electron temperature at the front surface of gold film calculated from the BTE, the EHD model, and theTTM. Evolution of (b) electron density fluctuation and (c) particle current density of electrons in the film calculated from the BTE and the EHD models at different delay time, respectively. The calculations use a laser fluence $I = 1\text{mJ/cm}^2$ and film thickness $L = 50\text{nm}$ as input parameters.

FIG. 3 (Color online) Time-dependent electron temperature at the front surface of the film calculated from the BTE, the EHD model, and the TTM after ultra-short pulse laser heating with different thicknesses and laser durations of (a) $L = 50\text{nm}$ and $\tau_p = 150\text{fs}$, (b) $L = 50\text{nm}$ and $\tau_p = 200\text{fs}$, (c) $L = 40\text{nm}$ and $\tau_p = 96\text{fs}$, (d) $L = 60\text{nm}$ and $\tau_p = 96\text{fs}$. The calculations use laser fluence $I = 1\text{mJ/cm}^2$ as input parameter.

FIG. 4 (Color online) Time-dependent electron temperature at front surface of the film calculated from the BTE, the EHD model, and the TTM for different laser fluences of 5mJ/cm$^2$ and 10mJ/cm$^2$. The calculations use laser duration $\tau_p = 96$fs and film thickness $L = 50$nm as input parameters.

FIG. 5 (Color online) Time-dependent electron and phonon temperature at the front surface calculated from the EHD model and the TTM for different film thicknesses of 25nm and 50nm for a high laser fluence $I = 100$mJ/cm$^2$ with a laser duration $\tau_p = 96$fs.

FIG. 6 (Color online) Comparison of the damage threshold for a $\tau_p = 96$fs laser pulse as a function of film thickness, calculated from the EHD model and the TTM.

Figure 1

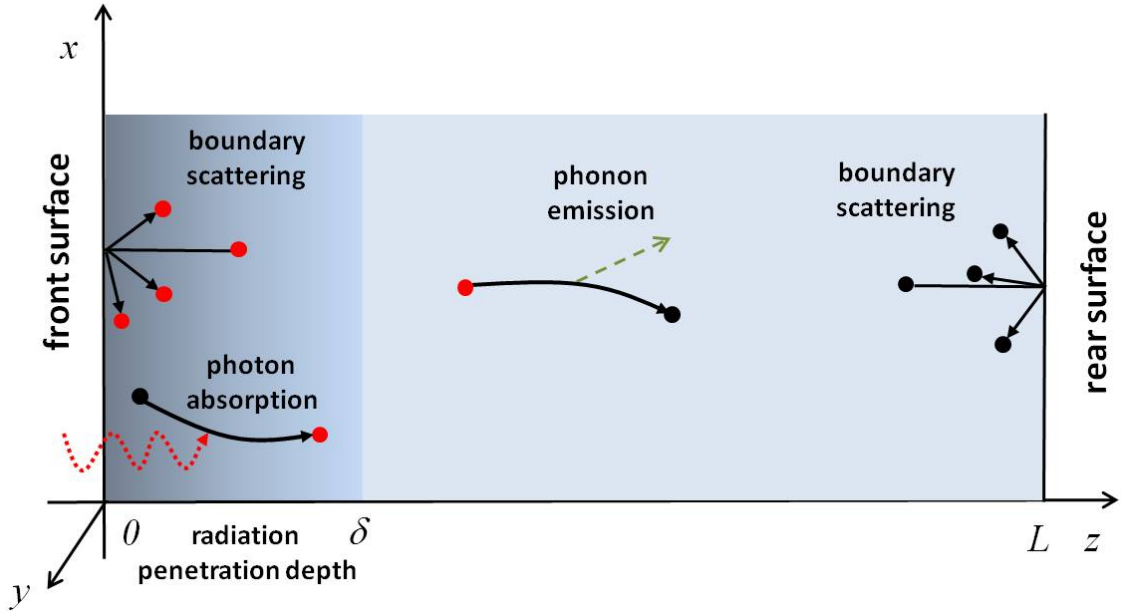

FIG. 1 (Color online) Schematic diagram of the coupled non-equilibrium electron and phonon transport in metal film after ultra-short pulse laser heating. The film thickness is $L$ and the surface of the metal film is in *x-y* plane. The low energy electrons (black dot) near the front surface are excited to be higher energy ones (red dot) after absorbing a photon. The absorption intensity exponentially decays with the radiation penetration depth $\delta$. The hot electrons release energy through two channels: transport to the rear surface and colliding with the lattice to emit phonons, i.e. EP scattering. When the electrons reach both the front and rear surfaces, they are scattered back randomly if a rough surface is assumed. In this work, diffusive reflection boundary conditions of electrons and phonons are used.

Figure 2

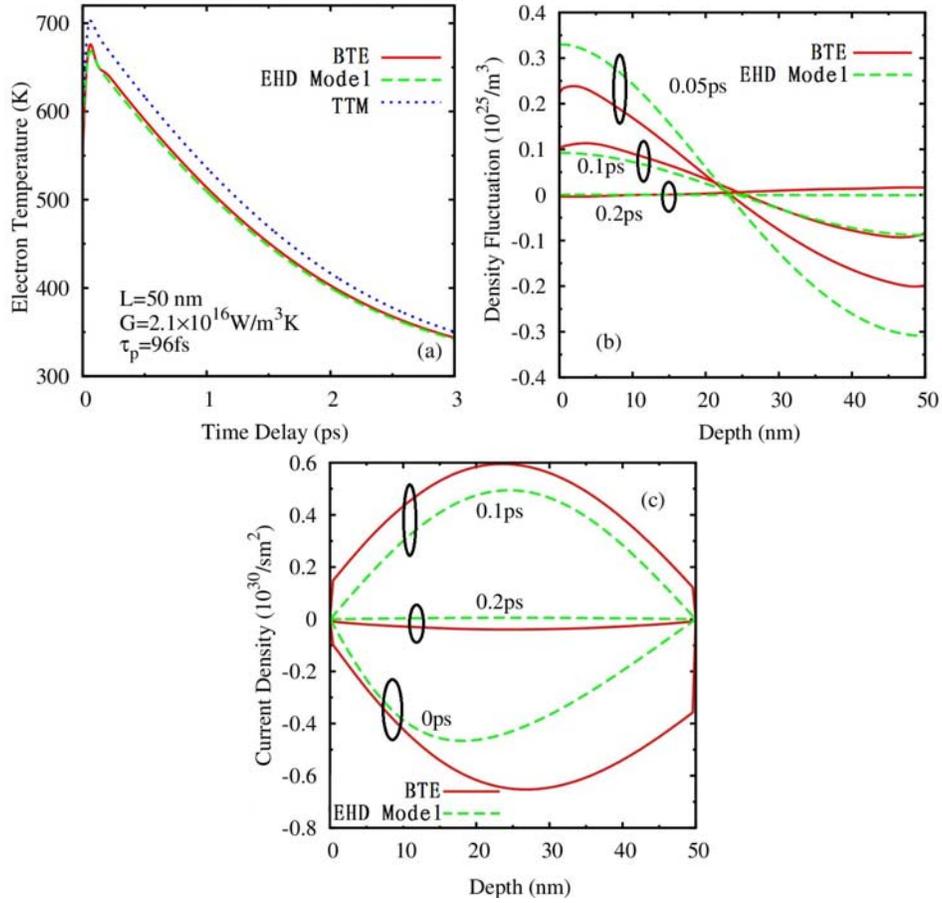

FIG. 2 (Color online) (a) Time-dependent electron temperature at the front surface of gold film calculated from the BTE, the EHD model, and the TTM. Evolution of (b) electron density fluctuation and (c) particle current density of electrons in the film calculated from the BTE and the EHD models at different delay time, respectively. The calculations use a laser fluence $I = 1\text{mJ/cm}^2$ and film thickness $L = 50\text{nm}$ as input parameters.

Figure 3

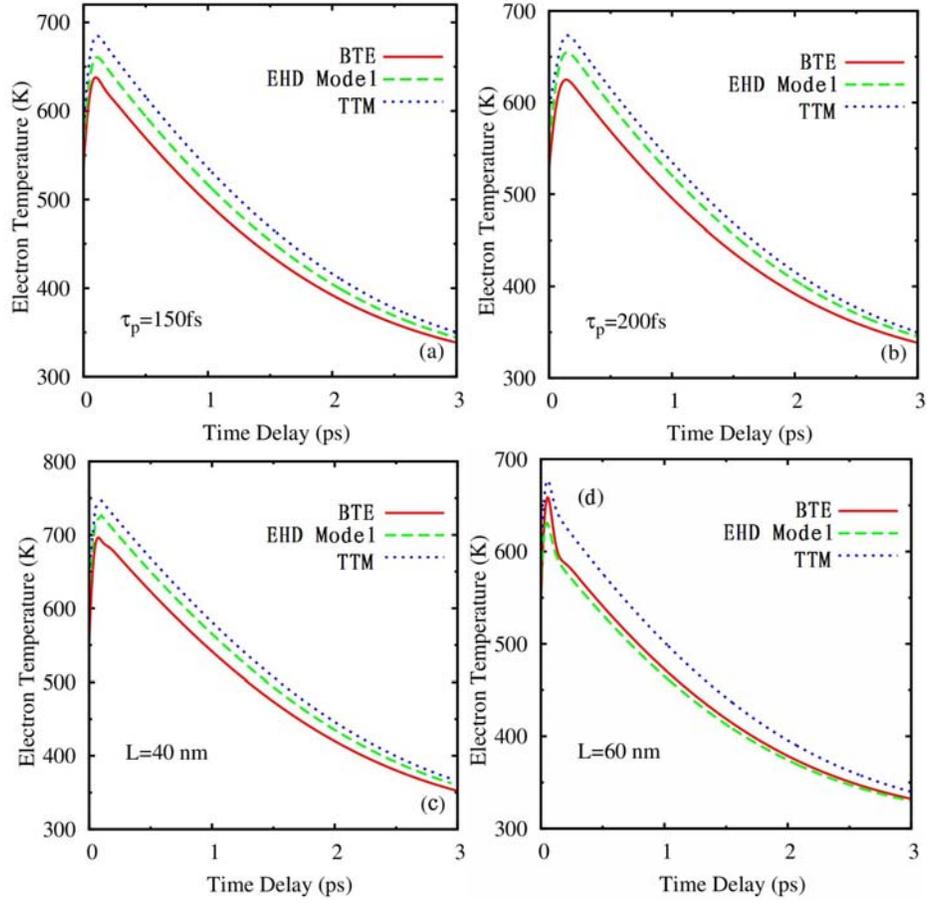

FIG. 3 (Color online) Time-dependent electron temperature at the front surface of the film calculated from the BTE, the EHD model, and the TTM after ultra-short pulse laser heating with different thicknesses and laser durations of (a) $L = 50\text{nm}$ and $\tau_p = 150\text{fs}$, (b) $L = 50\text{nm}$ and $\tau_p = 200\text{fs}$, (c) $L = 40\text{nm}$ and $\tau_p = 96\text{fs}$, (d) $L = 60\text{nm}$ and $\tau_p = 96\text{fs}$. The calculations use laser fluence $I = 1\text{mJ/cm}^2$ as input parameter.

Figure 4

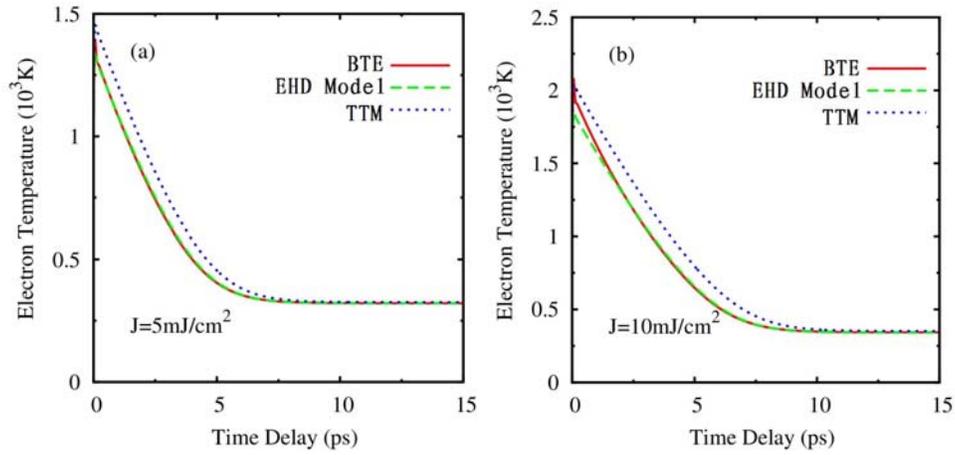

FIG. 4 (Color online) Time-dependent electron temperature at front surface of the film calculated from the BTE, the EHD model, and the TTM for different laser fluences of $5\text{mJ/cm}^2$ and $10\text{mJ/cm}^2$. The calculations use laser duration $\tau_p = 96\text{fs}$ and film thickness $L = 50\text{nm}$ as input parameters.

Figure 5

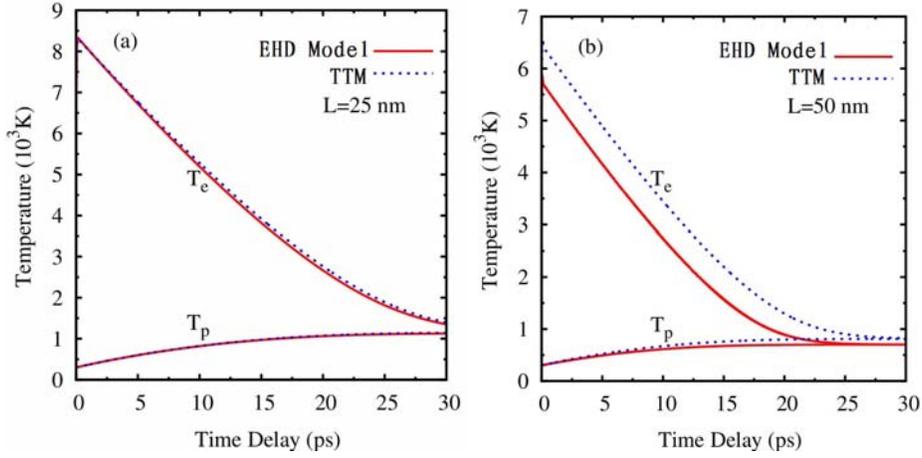

FIG. 5 (Color online) Comparison of time-dependent electron and phonon temperature at the front surface calculated from the EHD model and the TTM for different film thicknesses of 25nm and 50nm for a high laser fluence $I = 100$mJ/cm$^2$ with a laser duration $\tau_p = 96$fs.

Figure 6

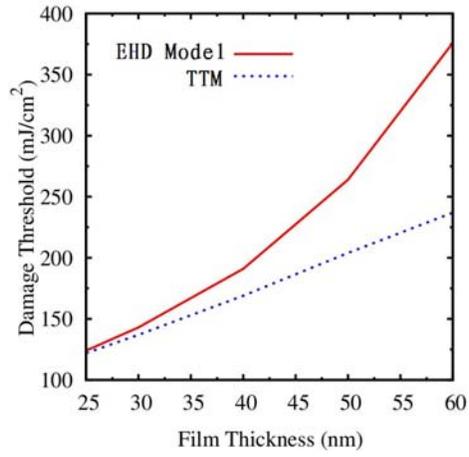

FIG. 6 (Color online) Comparison of the damage threshold for a $\tau_p = 96\text{fs}$ laser pulse as a function of film thickness, calculated from the EHD model and the TTM.